\documentclass[12pt,a4paper]{article}
\usepackage{amssymb,graphicx,graphics,epsfig}
\usepackage{lscape,graphics,amsmath}
\usepackage{latexsym,amsfonts}
\usepackage{axodraw}
\usepackage[english]{babel}

\newcounter{diags}
\newcounter{diags1}

\usepackage{epsfig}
\input epsf

\begin{document}
\begin{flushright}

\end{flushright}
%                          Title
\begin{center}
{\Large \bf The specificity of searches for $W^{\prime}$,
$Z^{\prime}$ and  $\gamma^{\prime}$  coming from extra dimensions}

\vspace{4mm}

%                      author/address
Edward E.~Boos, Viacheslav E.~Bunichev, Maxim A.~Perfilov, Mikhail
N.~Smolyakov,\\ Igor P.~Volobuev \\
\vspace{4mm}
 Skobeltsyn Institute of Nuclear Physics,
Moscow State University
\\ 119991 Moscow, Russia \\

\end{center}

\begin{abstract}
We discuss the specificity of searches for hypothetical
$W^{\prime}$, $Z^{\prime}$ and  $\gamma^{\prime}$ bosons at hadron
colliders in single top quark and $\mu^{+}\nu_{\mu}$ production
and Drell-Yan processes assuming these particles to be the
Kaluza-Klein excitations of the  gauge bosons of  the Standard
Model. In this case any process mediated by $W$ is also mediated
by the whole KK tower of its excitations, whereas to the processes
mediated by $Z$ and $\gamma$ there is not only a contribution from
their KK towers, but also from that of  the graviton. The
contributions of the towers above $W^{\prime}$, $Z^{\prime}$ and
$\gamma^{\prime}$ and above the first excitation of the graviton
are included with the help of effective four-fermion Lagrangians.
We compute the cross-sections of these processes taking into
account the contributions of the Standard Model gauge bosons, of
their first KK modes and of the corresponding KK towers and
discuss the impact of the interference between them. For
pp-collisions at the LHC with the center of mass energy
$14\,\textrm{TeV}$ we found specific changes of the distribution
tails due to the interference effects. Such a modification of
distribution tails is characteristic for the processes mediated by
particles coming from extra dimensions and should always be taken
into account when looking for them.
\end{abstract}

\section{Introduction}
Many theories and theoretical schemes for physics beyond the
Standard Model (SM) predict the existence of massive charged and
neutral vector particles in addition to the gauge bosons of the
SM. Such particles can arise either due to an extension of the SM
gauge group \cite{ex_gauge}, or as excitations of the SM gauge
bosons (see, for example \cite{ex_boson}); the lowest excitations
of $W$, $Z$ and $\gamma$ are usually called $W^{\prime}$,
$Z^{\prime}$ and $\gamma^{\prime}$.

The physical properties and the interactions of these particles
are different in different models. In some models they can have
couplings similar to those of the SM gauge bosons and can mediate
the same processes with SM particles. In paper \cite{Boos:2006xe}
it was noted that in this case  a nontrivial interference between
the contributions of $W$ and $W^{\prime}$ to various processes
could influence an experimental observation  of the latter in the
energy range close to its mass or the exclusion limits for it. In
particular, the negative interference resulted in observed weaker
exclusion limits in the case of the left-interacting compared to
the right-interacting $W^{\prime}$ \cite{Abazov:2011xs}. The
interference and its consequences  were also discussed in papers
\cite{Rizzo:2007xs,Accomando}.

If the additional vector bosons are found at the LHC, there arises
the problem of determining the theory beyond the SM, to which they
correspond. To solve this problem one has to study the specific
features of the additional vector bosons in different models. In
the present paper we will do it for the excitations of the SM
gauge bosons in models with universal extra dimensions (UED),
which have been widely discussed lately
\cite{ACD}--\cite{Agashe:2013fda}.

In such brane-world models, not only the gravity, but also
certain  fields of the SM  propagate either  in the flat bulk
 \cite{ACD, RIZ, MMN}, or in the Randall-Sundrum bulk
 \cite{Pomarol:1999ad}--\cite{Agashe:2013fda}.
It is very natural to assume that only the SM gauge fields may
propagate in the bulk since there is no consistent mechanism for
trapping them on the brane \cite{Rubakov:2001kp}. In contrast,
such a mechanism exists for the fermion fields
\cite{Rubakov:2001kp}, and therefore the fermions may be trapped
on a brane or localized in its neighborhood.

 Here we will consider a scenario, where only gauge fields live in
the   bulk of a stabilized brane-world model
\cite{Goldberger:1999uk,DeWolfe:1999cp,Boos:2005dc}, i.e., in the
bulk between the branes with the separation distance stabilized by
a scalar field and with a warped bulk metric different from that
of the Randall-Sundrum model \cite{Randall:1999ee}. Unlike the UED
models with the flat bulk \cite{ACD, RIZ, MMN}, such models give
rise to different wave functions for the fields of different
tensor type and, for this reason, do not necessarily lead to the
KK number conservation. Therefore, a production of single KK
states is possible in such a scenario. However, FCNC currents,
which are strongly suppressed by the present-day experimental
data, do not appear in this case since the neutral currents have
the same diagonal structure as in the SM.

In the case where only the SM gauge fields propagate in the
Randall-Sundrum bulk, the  masses of the  KK excitations of the SM
gauge bosons have to be approximately larger than
$20\,\textrm{TeV}$ in order not to contradict the EW precision
data  \cite{Davoudiasl:1999tf} . Such heavy states are obviously
out of the reach of the $14\,\textrm{TeV}$ LHC (one might  hope to
detect the states at the $33\,\textrm{TeV}$ LHC, if such a
collider is realized). However, in the stabilized brane-world
models, where the  warp factor is different from the exponential
of a linear function, as it is in the Randall-Sundrum model, the
couplings of KK modes to the SM fields might be significantly
different from those in the Randall-Sundrum model and, as a
result, lighter KK excitations of the SM fields may be allowed. A
study  of such stabilized brane-world models has been carried out
in papers \cite{Boos:2005dc,Boos:2004uc}, and it was found that
they may also solve the hierarchy problem of the gravitational
interaction, give rise to the masses of KK excitations in the
$\textrm{TeV}$ energy range, but the corresponding equations
cannot be solved exactly and should be studied numerically.

In order to  present better the physics of the processes mediated
by the KK excitations of gauge bosons, here we will not carry out
such calculations for a specific stabilized brane-world model, but
rather give a qualitative description of the phenomena taking for
the masses  values close to the proposed benchmark Snowmass 2013
parameter points \cite{Agashe:2013fda}, which have been chosen to
cover the energy range of the LHC experiments at
$14\,\textrm{TeV}$.  Such a choice of the KK masses seems to be
very useful for comparing our results with experimental data at
the $14\,\textrm{TeV}$ LHC. Though in our case, unlike in paper
\cite{Agashe:2013fda}, the SM fermions are supposed to be
localized on a brane and the coupling constants of the KK modes
are assumed to be the same, as in the SM, it is not difficult to
reproduce our results with different values of coupling constants.

A study along these lines of collider processes mediated by the
excitations of the SM electroweak gauge bosons propagating in the
bulk of a stabilized brane-world model was performed in paper
\cite{Boos:2011ib} exposing the role of the interference. Here we
will briefly recall the results of this paper and elaborate them
for a number of  processes mediated by both charged and neutral
gauge bosons and their KK excitations, taking into account also
the contribution of the graviton resonances in the latter case. We
consider both the processes with charged and neutral KK
excitations, because all of them   should manifest simultaneously
in the LHC experiments.

The processes mediated by the neutral particles including the
graviton excitations have been discussed in
\cite{Davoudiasl:2000wi} within the framework of the unstabilized
Randall-Sundrum model for the masses of the first KK excitations
lying below $2\,\textrm{TeV}$, this energy range having been
studied by the LHC nowadays  \cite{Chatrchyan:2013qha}. The
presence of a destructive interference between $\gamma'$ and $Z'$
resonances in models with large extra dimensions was already noted
in paper \cite{Accomando:1999sj} without taking into account the
contributions of the higher excitations. Here we extend the
analysis of the  processes with the intermediate neutral particles
to a more general setting  of stabilized Randall-Sundrum models
and to a larger, not yet excluded, energy range taking into
account contributions of the KK towers and all the interferences.

\section{Effective interactions}
The characteristic feature of theories with compact extra
dimensions is the presence of towers of Kaluza-Klein excitations
of the bulk fields, all the excitations of a bulk field having the
same type of coupling to the fields of the SM. If we consider
these theories for the energy or momentum transfer much smaller,
than the masses of the KK-excitations, we can pass to the
effective ``low-energy" theory, which can be obtained by the
standard procedure. Namely, we have to drop the momentum
dependence in the propagators of the heavy modes and to integrate
them out in the functional integral built with the original
action. In the case, where only the gravity propagates in the bulk
of a stabilized brane-world model, the resulting Lagrangian turns
out to be \cite{BBSV}
\begin{eqnarray}\label{tensor_intCRS}
L_{eff}&=&\frac{C}{\Lambda_{\pi}^{2}M_{1}^{2}}
T^{\mu\nu} \tilde \Delta_{\mu\nu, \rho\sigma}T^{ \rho\sigma},\\
\tilde \Delta_{\mu\nu, \rho\sigma} &=&
\frac{1}{2}\eta_{\mu\rho}\eta_{\nu\sigma}  +
\frac{1}{2}\eta_{\mu\sigma}\eta_{\nu\rho}-\left(\frac{1}{3}-\frac{\delta}{2}\right)
\eta_{\mu\nu}\eta_{\rho\sigma}.
\end{eqnarray}
Here $T^{\mu\nu}$ stands for the energy-momentum tensor of the SM
fields, $M_1$ is the mass of the first tensor resonance,
$\Lambda_\pi$ is its (inverse) coupling constant to the SM fields,
the constant $\delta$ takes into account the contribution of the
scalar modes, the dimensionless constant $C$ is defined by the
geometry of the model and can be computed numerically. In
particular, in paper \cite{BBSV} it was found to be equal
approximately to $1.8$ in the stabilized Randall-Sundrum model. In
this paper it was also noted that if the center of mass energy is
close to the mass of the first resonance, its contribution should
be taken into account exactly, whereas the rest of the tower can
still be approximated by Lagrangian (\ref{tensor_intCRS}) with the
effective coupling constant ${0.8}/({\Lambda_{\pi}^{2}M_{1}^{2}})$
instead of ${1.8}/({\Lambda_{\pi}^{2}M_{1}^{2}})$.

A contact interaction Lagrangian  can be obtained in the same way
for the  interactions mediated by the $SU(2) \times U(1)$ bulk
gauge fields  \cite{Boos:2011ib}. These fields are described in
the bulk by vector potentials $W_M$ and $B_M$, which give rise to
four-dimensional vector and scalar fields. The latter are in the
trivial and in the adjoint representations of $SU(2)$ and cannot
break $SU(2) \times U(1)$ to $ U(1)_{em}$, as it is necessary in
the SM. For this reason, we assume that the gauge symmetry is
broken in the standard way by the Higgs field on the brane. It is
useful to pass in the standard way to  the charged vector fields
$W^{\pm}_\mu$ and  the physical  neutral vector  fields $Z_{\mu}$
and  $ A_{\mu}$. After the spontaneous symmetry breaking  the
neutral component of the brane Higgs field acquires the vacuum
value $v/\sqrt{2}$, and there arises a quadratic  interaction of
the KK modes of the vector fields proportional to the product of
the values of the wave functions of the modes on the brane and to
$M_W^2$ for charged fields and to $M_Z^2$ for neutral fields. Due
to this interaction the zero modes of the fields $W^{\pm}_\mu$ and
$Z_{\mu}$  acquire masses and the KK modes are no longer the mass
eigenstates; the latter are now superpositions of the modes
\cite{Muck:2002af}. But if the mass scale generated by the Higgs
field  is much smaller, than the mass of the first KK excitation
-- and it is exactly the scenario we are studying -- this mixing
of KK modes can be neglected \cite{Muck:2002af}. In this
approximation the masses of the excitations of the $W$ and $Z$
bosons, as well as those of the photon, should be treated as
nearly equal. This is due to the fact that all vector fields
satisfy the same equation  of motion in the bulk, and for this
reason the KK masses of their excitations are the same, which is
also true for more complicated models \cite{Frere:2003ye}. It is
also worth  noting that a model-independent analysis based only on
the properties of the SM gauge group representations carried by
the fields of heavy charged and neutral bosons coupled to leptons
also leads to a similar degeneracy of their masses
\cite{deBlas:2012qp}.

The interaction vertices of the KK modes and the fields of the SM
are the same as those of the zero modes. Integrating out the heavy
modes, we again arrive at an effective Lagrangian for the
interaction of the SM fields due to the excitations of the
$SU(2)\times U(1)$ gauge bosons
\begin{equation}\label{effl_V}
L_{eff}= \frac{g^2 }{M^2}\left( C_W J^{+\mu }J^{-}_{\mu } + C_Z
J^{(0)\mu }J^{(0)}_{\mu } + C_A J^{\mu }_{em}J_{em\,\mu }\right),
\end{equation}
$M$ being the energy scale of the model, $g$ denoting the $SU(2)$
gauge coupling constant and $J^{+}_{\mu }, J^{(0) }_\mu, J^{\mu
}_{em} $ being the SM weak charged, weak neutral and
electromagnetic currents. The constants $C_W, C_Z, C_A$ are again
model dependent and can be estimated only in a specific model.

In the case of  the $SU(2) \times U(1)$  gauge fields  in the
5-dimensional bulk, one can pass to the axial gauge, where the
components corresponding to the extra dimension are equal to zero.
Thus, the gauge sector of the original 5-dimensional theory does
not add scalar fields to the effective four-dimensional theory. In
any brane-world model, the mass spectrum of gauge fields is
defined by  a Sturm-Liouville eigenvalue problem with Neumann
boundary conditions, the wave functions $w_n(y)$ of the fields
$A^n_\mu(x)$ with definite masses being its solutions. Due to this
fact the wave function of the massless zero mode is constant in
the extra dimension, which guarantees the universality of its
coupling constant \cite{Rubakov:2001kp}.  In particular, in the
Randall-Sundrum model \cite{Randall:1999ee}, the masses of the
lower excitations of vector fields are approximately given by the
zeros of the Bessel function $J_0(M_n/k)$ \cite{Davoudiasl:1999tf}
($k$ being the energy scale of the model), whereas those of the
tensor fields are approximately given by the corresponding zeros
of the Bessel function $J_1(M_n/k)$ \cite{Davoudiasl:1999jd},
which are always larger \cite{BE}. Thus, if the  energy scale of
the model is in the \textrm{TeV} energy range, the masses of the
modes appear to be also in the \textrm{TeV} energy range. In fact,
this is valid in stabilized  brane-world models as well. We have
already noted in the Introduction that the phenomenological
constraints on the masses of the gauge boson excitations  in paper
\cite{Davoudiasl:1999tf} do not apply, in general, to stabilized
models because   the warp factor of  a stabilized brane-world
model differs from the exponential of a linear function in the
unstabilized Randall-Sundrum model.

We will be interested in the case where the masses of the modes
and the mass gaps between them are quite large, say, of the order
of a few \textrm{TeV}. In particular, in the UED models with the
energy scale $1\,\textrm{TeV}$, the mass of the lowest gauge boson
excitations and that of the first graviton excitation are (almost)
the same, whereas in the unstabilized Randall-Sundrum model they
are $2.4\,\textrm{TeV}$  for the lowest gauge boson excitations
and $3.8\,\textrm{TeV}$ for the lowest graviton excitation, their
ratio being approximately equal to 1.6 and tending to 1 for the
higher excitations. Thus, we expect this ratio to be of the order
of unity, and will use the value 1.5 in our subsequent
calculations.

Below we will consider some processes with the Kaluza-Klein
electroweak gauge bosons at the energies accessible at the LHC
supposing that the masses of $W'$, $Z'$ and $\gamma'$ are within
this energy range. Similar to the approach of paper \cite{BBSV} we
exactly take into account the contributions of the first
Kaluza-Klein modes, whereas the contributions of all the other
modes are taken into account by means of the effective contact
interaction (\ref{effl_V}), from which  the  contribution of the
first mode is subtracted.

Symbolic and numerical computations, including simulations of the
SM background for the LHC, have been performed by means of the
CompHEP package \cite{comphep}, into which the corresponding
Feynman rules for the new states and interactions have been
implemented.

\section{Two-body processes mediated by excitations of gauge bosons}

First we consider the simpler case of the $W$ boson and its KK
tower. The coupling constants of its excitations and their masses
essentially depend on the fundamental parameters of a stabilized
brane-world model, which is also true for the excitations of other
particles to be discussed below. In particular, in paper
\cite{BBSV} the masses of the graviton excitations were explicitly
calculated in terms of the fundamental parameters, which turned
out to be a rather complicated task.

As we mentioned in the Introduction, for our study we choose
values for the excitations masses in the energy range of the LHC,
but not yet excluded, and close to the Snowmass 2013 benchmarks
\cite{Agashe:2013fda}. We also note that for all the taken
parameter points the width of an excitation is not larger that its
mass as it should be.

Then the effective interaction  Lagrangian of the $W$ boson  KK
tower in the energy range close to the mass of the $W^{\prime}$
boson looks like
\begin{equation}\label{W_KK}
L_{eff\_W\_KK} =  \frac{g_1 }{\sqrt{2}}( J^{+\mu }W^{\prime-}_\mu
+ J^{-\mu }W^{\prime+}_\mu) - \frac{g_1^2 }{2
M^2_{W^{\prime}\_sum}} J^{+\mu }J^{-}_{\mu },
\end{equation}
where $g_1$ is the coupling constant of $W^{\prime}$ to the weak
charged current $J^{+\mu } $ and $M_{W^{\prime}\_sum}$ is the mass
parameter taking into account the contribution of the KK tower
above $W^{\prime}$. Thus, this Lagrangian has three free
parameters, including the $W^{\prime}$ mass.

Under these assumptions we will study the processes $pp \to t \bar
b + X$  and $pp \to \mu^+ \nu_\mu + X$ at the LHC.  We start with
the single top production. It  occurs due to the weak process $u
\bar d \to t \bar b$, which is mediated by the $W$ boson and its
KK tower. In our approximation the amplitude of the process can be
represented by the diagrams

\vspace*{1cm}
\begin{picture}(93,81)(0,0)
\ArrowLine(13.5,78.0)(40.5,64.5) \Text(12.5,78.0)[r]{$u$}
\ArrowLine(40.5,64.5)(13.5,51.0) \Text(12.5,51.0)[r]{$\bar d$}
\Photon(40.5,64.5)(67.5,64.5){3}{3.0} \Text(54.8,70.5)[b]{$W$}
\ArrowLine(94.5,78.0)(67.5,64.5) \Text(96.5,78.0)[l]{$\bar b$}
\ArrowLine(67.5,64.5)(94.5,51.0) \Text(96.5,51.0)[l]{$t$}
\Text(111.8,60.5)[b]{\begin{Large}$+$\end{Large}}
%\Text(46,0)[b] {diagr.1}
\end{picture} \
\vspace{5mm} {} \qquad\allowbreak
%  diagram # 2
\begin{picture}(93,81)(0,0)
\ArrowLine(13.5,78.0)(40.5,64.5) \Text(12.5,78.0)[r]{$u$}
\ArrowLine(40.5,64.5)(13.5,51.0) \Text(12.5,51.0)[r]{$\bar d$}
%\DashArrowLine(40.5,64.5)(67.5,64.5){3.0}
%\DashArrowLine(40.5,64.5)(67.5,64.5){3.0}
\Photon(40.5,64.5)(67.5,64.5){3}{3.0}
\Text(54.8,70.5)[b]{$W^{\prime}$} \ArrowLine(94.5,78.0)(67.5,64.5)
\Text(95.5,78.0)[l]{$\bar b$} \ArrowLine(67.5,64.5)(94.5,51.0)
\Text(95.5,51.0)[l]{$t$}
\Text(111.8,60.5)[b]{\begin{Large}$+$\end{Large}}
%\Text(46,0)[b] {diagr.2}
\end{picture} \
\vspace{5mm} {} \qquad\allowbreak
%  diagram # 3
\begin{picture}(93,81)(0,0)
\ArrowLine(13.5,78.0)(30.5,64.5) \Text(12.5,78.0)[r]{$u$}
\ArrowLine(30.5,64.5)(13.5,51.0) \Text(12.5,51.0)[r]{$\bar d$}
%\DashArrowLine(40.5,64.5)(67.5,64.5){3.0}
%\Text(54.8,67.5)[b]{$A+$}
%\ArrowLine(94.5,78.0)(67.5,64.5)
\Text(23.5,76.0)[l]{{\scriptsize  CW}}
\Text(27.0,64.0)[l]{\begin{large}$ \bullet$\end{large}}
\ArrowLine(47.5,78.0)(30.5,64.5) \Text(49.0,78.0)[l]{$\bar b$}
\ArrowLine(30.5,64.5)(47.5,51.0) \Text(49.0,51.0)[l]{$t$}
%\Text(184.8,60.5)[b]{(14)}
%\Text(46,0)[b] {diagr.3}
\Text(185.8,60.5)[b]{\addtocounter{equation}{1}(\arabic{equation})}
\setcounter{diags}{\value{equation}}\label{diags1}
\end{picture}

\vspace*{-2cm} \noindent The contact term {CW} is, in fact,
Fermi's interaction with the  coupling constant $g_1^2/(
2M^2_{W^{\prime}\_sum})$.  Explicit  calculations in UED models
with flat extra dimension and in  certain stabilized brane-world
models show that this effective mass is just a little larger than
that of $W^{\prime}$.\footnote{We recall once again that the
effective mass is a way to parameterize the sum of  the
contributions of the KK tower above the first resonance. This
point is also discussed in sections II and III of paper
\cite{BBSV}.} In the present paper we will take the value $ 1.4
M_{W^{\prime}}$, previously used in \cite{Boos:2011ib}, as the
effective mass of the KK tower above $W^{\prime}$ and will apply
the same relation to the excitations of the other gauge bosons.

For simplicity we will also assume that the coupling constants for
all the excitations of the SM gauge bosons are the same as those
of the gauge bosons themselves.  Then the amplitude corresponding
to  diagrams (\arabic{equation}) is equal to
\begin{equation}\label{amp}
\qquad \frac{g^2}{2} (\bar d \gamma_\mu (1-\gamma_5) u )( \bar t
\gamma^\mu (1-\gamma_5) b)
\left(\frac{1}{p^2-M_{W}^2}+\frac{1}{p^2-
M_{W'}^2}-\frac{1}{M^2_{W^{\prime}\_sum}}\right).
\end{equation}
 The last term in the brackets effectively takes into account the contribution of
the sum of the one-boson-exchange diagrams with all the modes
above the $W'$ boson. This structure of the amplitude manifests
the origin of the interference between the contributions of
different diagrams.

\begin{figure*}[!h!]
\begin{center}

\begin{minipage}[t]{.45\linewidth}
%\begin{minipage}[t][2cm][t]{0.4\textwidth}
\centering
\includegraphics[width=80mm,height=70mm]{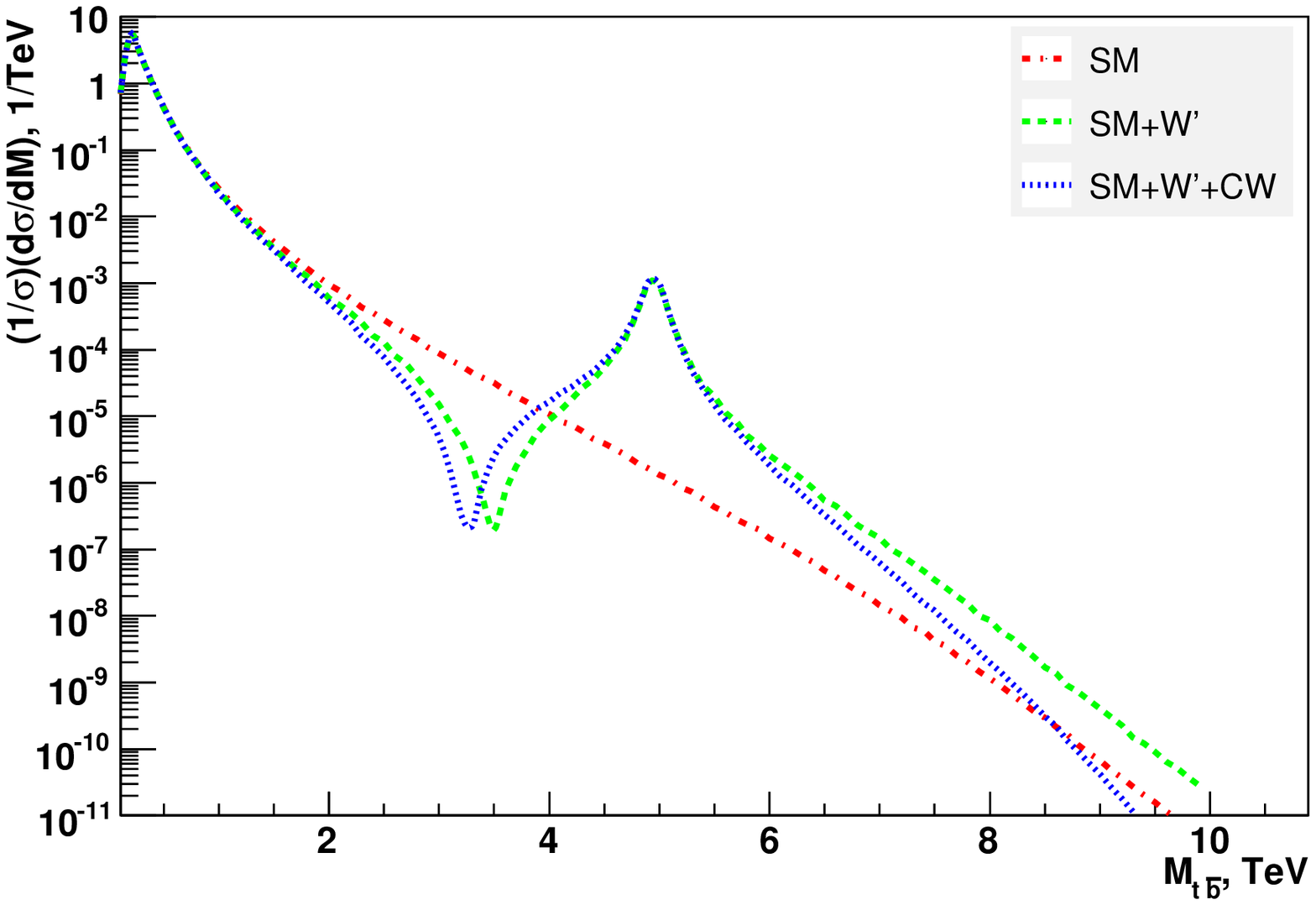}
\caption{\label{pic1_int} \footnotesize Invariant mass
distribution for the single top production at the LHC with the
center of mass energy $14~TeV$ for \,$M_{W'} = 5\,\textrm{TeV},\,
\Gamma_{W'} = 0.17\,\textrm{TeV}$ with and without the
contribution of the $W'$ KK tower. }
\end{minipage}
\hfill
\begin{minipage}[t]{.45\linewidth}
\centering
\includegraphics[width=80mm,height=70mm]{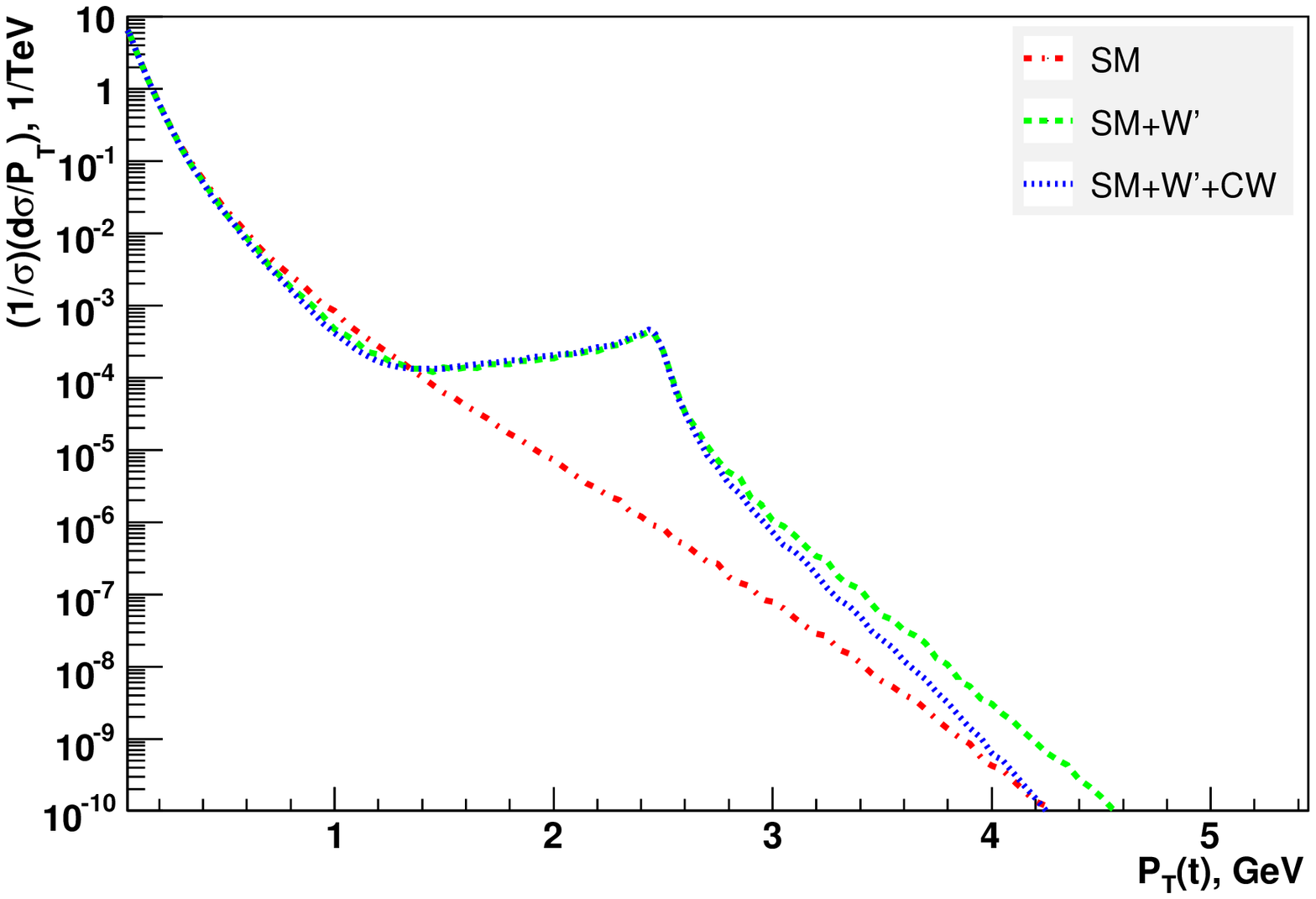}
\caption{ \label{pic2_int} \footnotesize $P_{T}$ distribution for
the single top production at the LHC with the center of mass
energy $14~TeV$ for $M_{W'} = 5\,\textrm{TeV},\, \Gamma_{W'} =
0.17\,\textrm{TeV}$ with and without the contribution of the $W'$
KK tower.}
\end{minipage}

\end{center}
\end{figure*}

The cross-sections of this process can be obtained by calculating
the corresponding partonic cross-sections and integrating them
with the parton distribution functions, and  in so doing we
neglect the  light quark masses. In particular, for $M_{W'} =
5\,\textrm{TeV}$, the corresponding distributions in the invariant
mass of the $t \bar b$  pair  and in the transverse momentum of
the top quark have been calculated for the first SM diagram only,
for the sum of the first two diagrams ($SM + W^{\prime} $), and
for all three diagrams ($SM + W^{\prime}+ CW $) and presented in
figures \ref{pic1_int} and \ref{pic2_int}.

%=========================================================================
%=============================================================================

These figures clearly show that the interference with the
contribution of the rest of the KK tower changes the curves
significantly (similar curves  were obtained in papers
\cite{Boos:2006xe} and \cite{Boos:2011ib} for the masses of the
$W^{\prime}$ boson $1\,\textrm{TeV}$ and $2\,\textrm{TeV}$).

\begin{figure*}[!h!]
\begin{center}

\begin{minipage}[t]{.45\linewidth}
%\begin{minipage}[t][2cm][t]{0.4\textwidth}
\centering
\includegraphics[width=80mm,height=70mm]{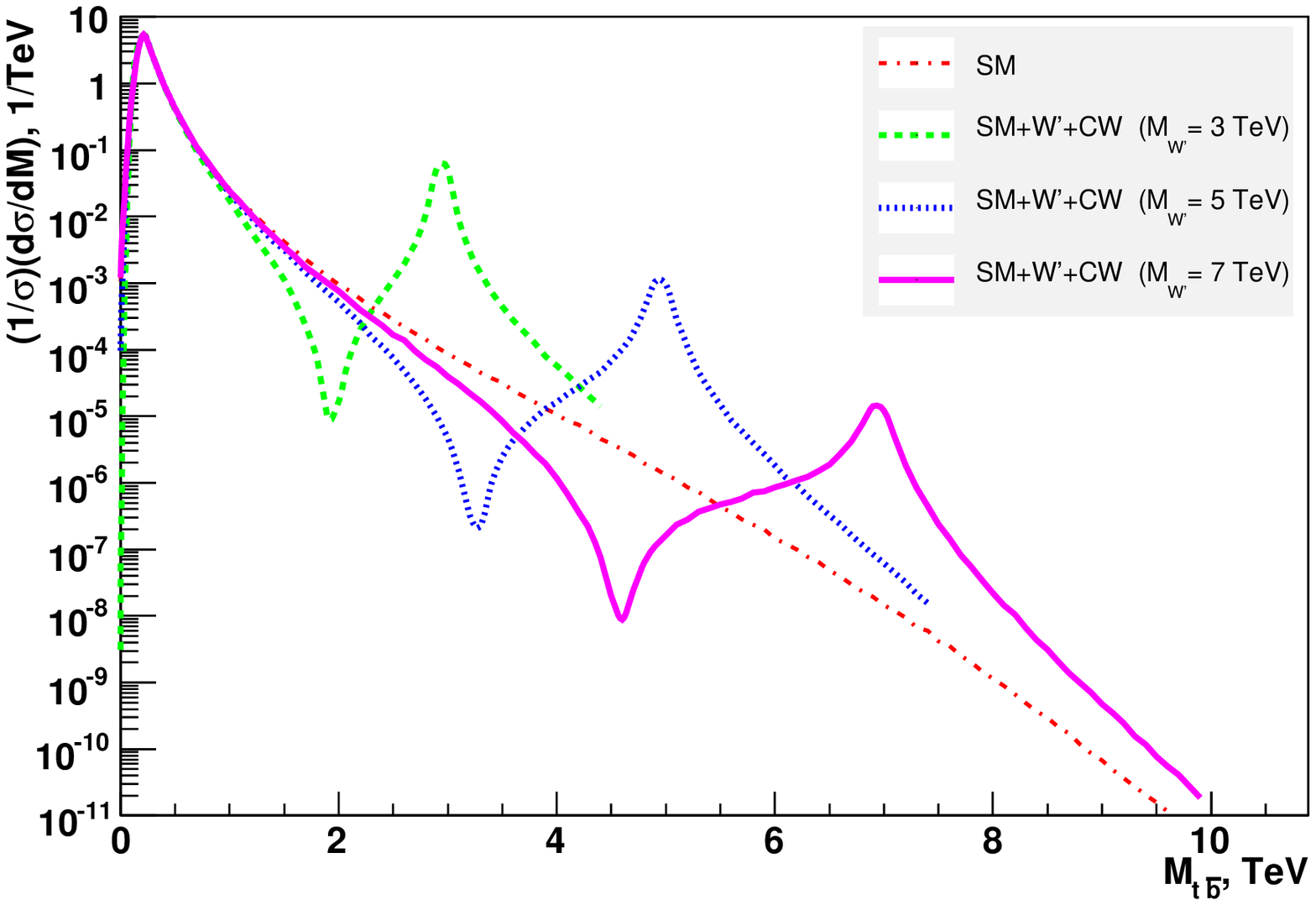}
\caption{\label{pic1} \footnotesize Invariant mass distribution for the single
top production at the LHC with the center of mass energy 14~TeV
for three different values of $M_{W'}$ mass:  $M_{W'} =
3\,\textrm{TeV},\, \Gamma_{W'} = 0.10 \,\textrm{TeV}$; \,$M_{W'} =
5\,\textrm{TeV},\, \Gamma_{W'} =  0.17\,\textrm{TeV}$; \, $M_{W'}
= 7\,\textrm{TeV}, \, \Gamma_{W'} = 0.23\,\textrm{TeV}$. }
\end{minipage}
\hfill
\begin{minipage}[t]{.45\linewidth}
\centering
\includegraphics[width=80mm,height=70mm]{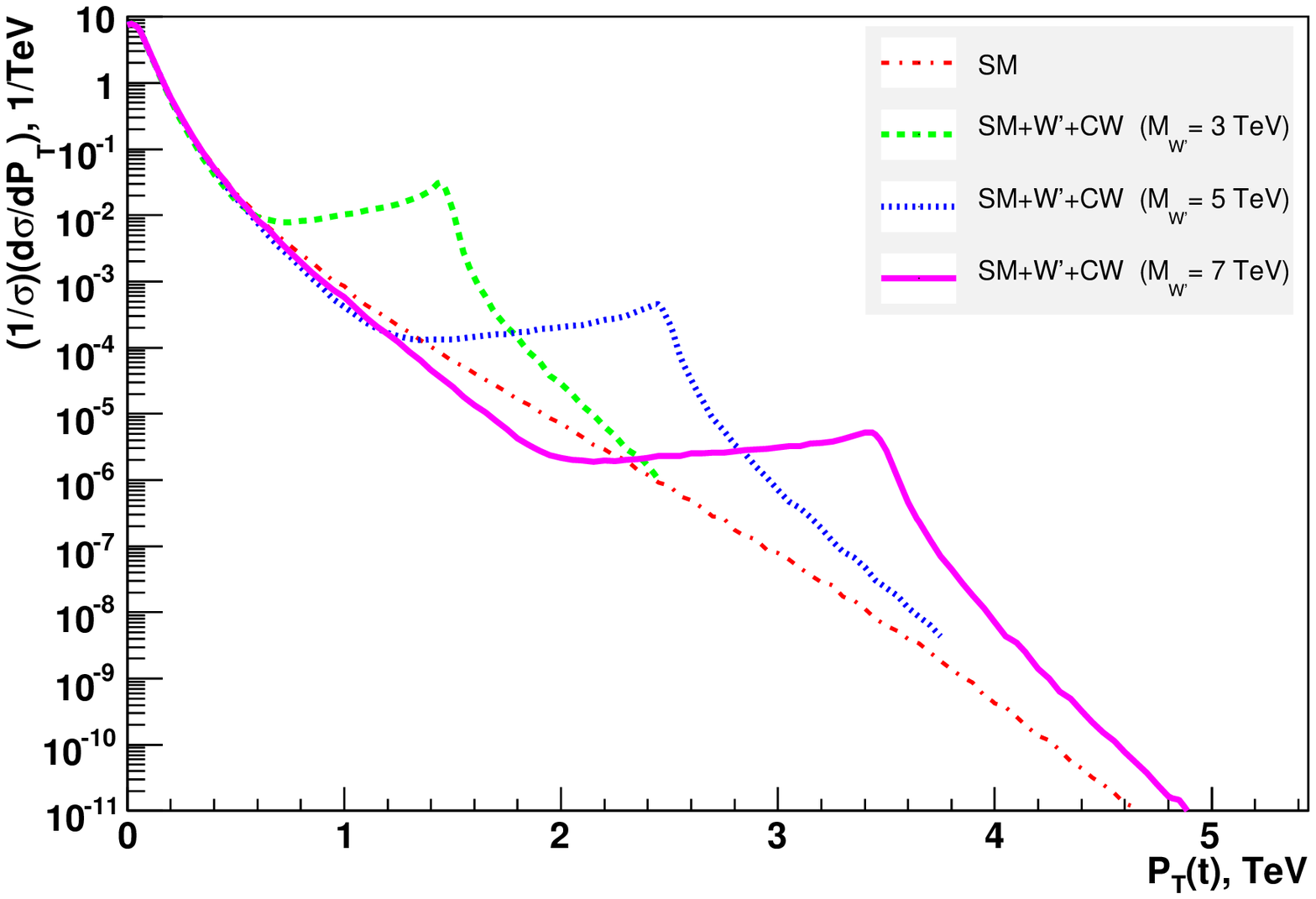}
\caption{ \label{pic2} \footnotesize $P_{T}$ distribution for the single top
production at the LHC with the center of mass energy 14~TeV for
three different values of $M_{W'}$ mass:  $M_{W'} =
3\,\textrm{TeV},\, \Gamma_{W'} = 0.10 \,\textrm{TeV}$; \,$M_{W'} =
5\,\textrm{TeV},\, \Gamma_{W'} =  0.17\,\textrm{TeV}$; \, $M_{W'}
= 7\,\textrm{TeV}, \, \Gamma_{W'} = 0.23\,\textrm{TeV}$. }
\end{minipage}

\end{center}
\end{figure*}

In order to study the  dependence of these results on the $W'$
mass, here we  calculated these distributions with the complete
set of diagrams (\arabic{diags}) for two more values of
$W^{\prime}$ mass, $M_{W'} = 3\,\textrm{TeV}$,  \, $M_{W'} =
7\,\textrm{TeV}$,  which also belong to the energy range not yet
studied at the LHC. For these values of the $W'$ mass its widths
have been found to be  $0.10 \,\textrm{TeV}$ and $
0.23\,\textrm{TeV}$. The results, together with those for
$W^{\prime}$ mass $5 \,\textrm{TeV}$, are presented in figures
\ref{pic1} and \ref{pic2} so that for each curve only that energy
range is shown where the contribution of the higher excitations
can be approximated by the contact term $CW$.  Our estimates show
that, if the coupling of the $W'$ boson to the SM fields is of the
same order, as that of the $W$ boson, the interference effects due
to the contribution of  the rest of the KK tower are, in
principle, observable for the $W'$ boson mass as large as
$7\,\textrm{TeV}$.

%=========================================================================
%=============================================================================

\begin{figure*}[!h!]
\begin{center}

\begin{minipage}[t]{.45\linewidth}
%\begin{minipage}[t][2cm][t]{0.4\textwidth}
\centering
\includegraphics[width=80mm,height=70mm]{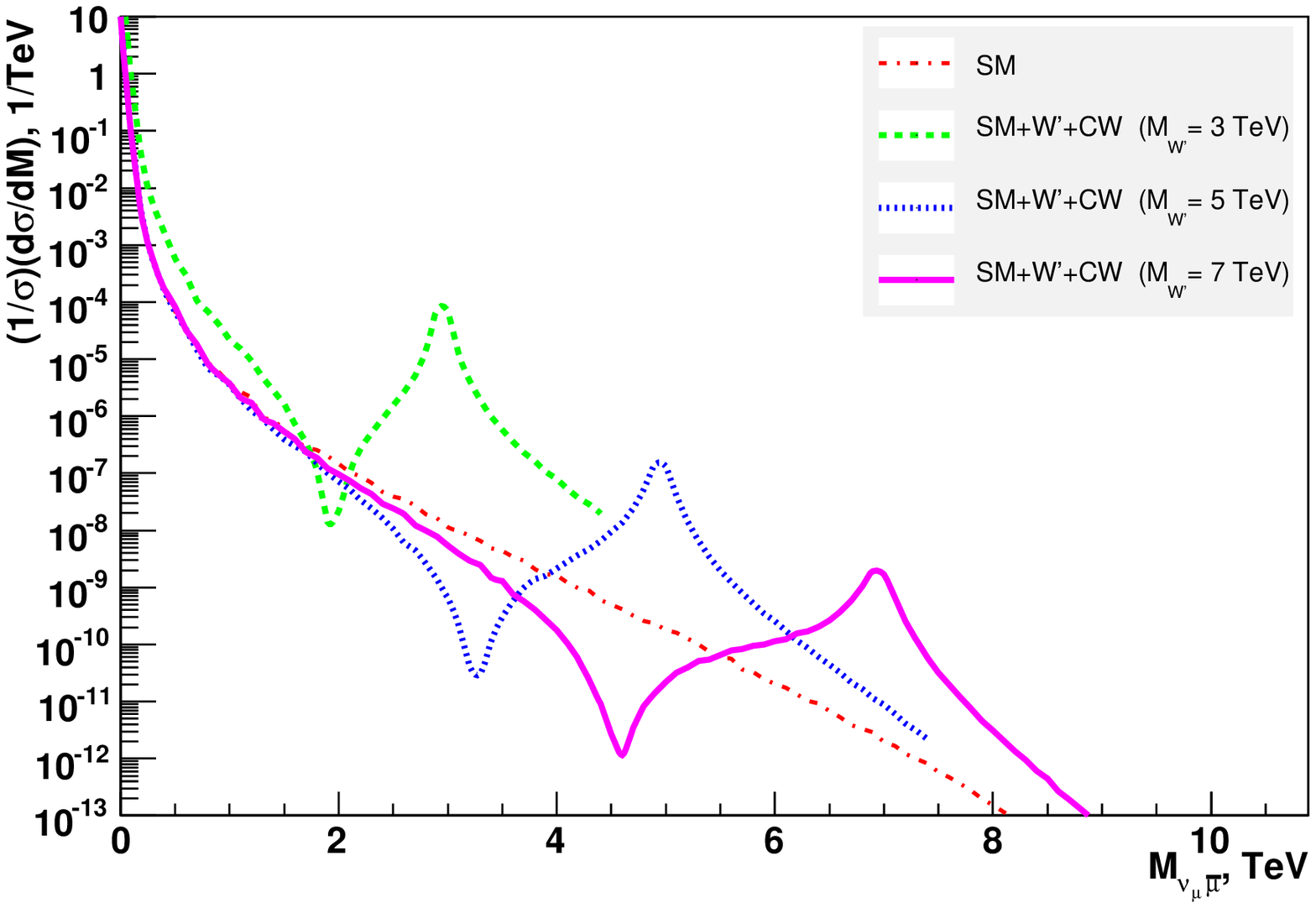}
\caption{\label{pic1m} \footnotesize Invariant mass distribution
for the $\mu^+$ production at the LHC with the center of mass
energy 14~TeV for three different values of $M_{W'}$ mass: $M_{W'}
= 3\,\textrm{TeV},\, \Gamma_{W'} = 0.10 \,\textrm{TeV}$; \,$M_{W'}
= 5\,\textrm{TeV},\, \Gamma_{W'} =  0.17\,\textrm{TeV}$; \,
$M_{W'} = 7\,\textrm{TeV}, \, \Gamma_{W'} = 0.23\,\textrm{TeV}$. }
\end{minipage}
\hfill
\begin{minipage}[t]{.45\linewidth}
\centering
\includegraphics[width=80mm,height=70mm]{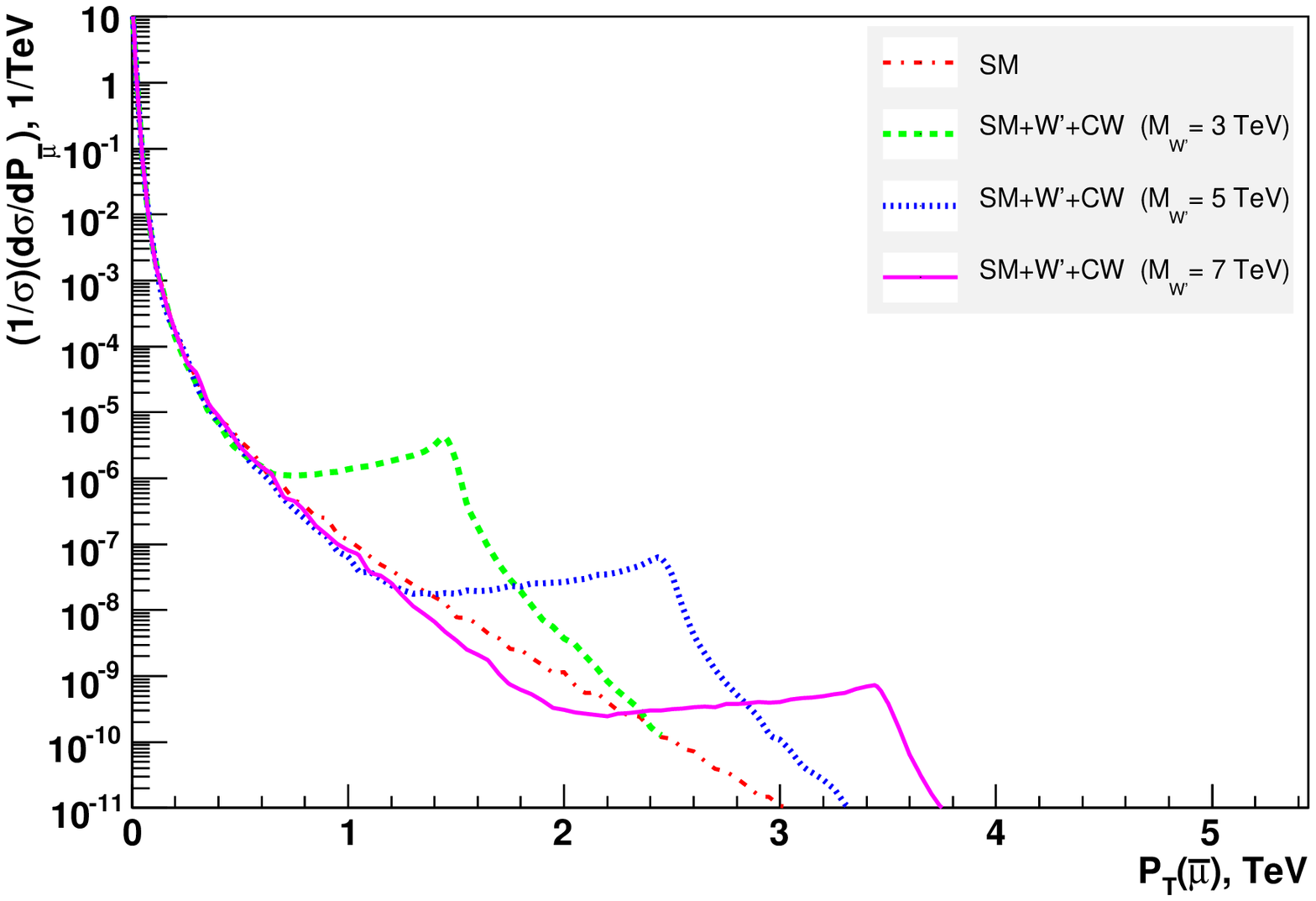}
\caption{ \label{pic2m} \footnotesize $P_{T}$ distribution for the
$\mu^+$ production at the LHC with the center of mass energy
14~TeV for three different values of $M_{W'}$ mass:  $M_{W'} =
3\,\textrm{TeV},\, \Gamma_{W'} = 0.10 \,\textrm{TeV}$; \,$M_{W'} =
5\,\textrm{TeV},\, \Gamma_{W'} =  0.17\,\textrm{TeV}$; \, $M_{W'}
= 7\,\textrm{TeV}, \, \Gamma_{W'} = 0.23\,\textrm{TeV}$. }
\end{minipage}

\end{center}
\end{figure*}

The process $pp \to \mu^+ \nu_\mu + X$ can be treated in the same
way. The corresponding diagrams can be easily obtained from
diagrams (5) by replacing the top quark by the neutrino and the
$\bar b$ quark by the positive muon. The cross-sections of this
process have been calculated for the same values of  the  mass of
the $W^\prime$ boson and the same characteristics of its KK tower,
the resulting plots being presented in figures \ref{pic1m} and
\ref{pic2m}. We see that the plots  look very much like those for
the top quark production.

%=========================================================================
%=============================================================================

The case of the $Z^{\prime}$ boson and $\gamma^{\prime}$ turns out
to be more complicated, because in theories with extra dimensions
the KK graviton $gr^{\prime}$ and its tower also contribute to all
the processes mediated by the neutral vector boson. (To be
precise, there is one more tower that contributes to all these
processes, namely that of the scalar radion. But the contribution
of the scalar modes is suppressed by the factor $(m_q/M)^2$, $m_q$
being the mass of a first generation quark  and $M$ being the
fundamental energy scale of the order of several \textrm{TeV}
\cite{BBSV}. For this reason the contribution of the scalar modes
is negligible and we discard it completely.) An example of
processes mediated by $Z^{\prime}$, $\gamma^{\prime}$ and the
corresponding KK towers is the Drell-Yan process $pp \to
\mu^+\mu^- + X$.

This process has been already studied at the LHC for the center of
mass energies $7\,\textrm{TeV}$ and $8\,\textrm{TeV}$ with the
result that  heavy narrow neutral resonances decaying to muon or
electron pairs should be heavier than  $2.6\,\textrm{TeV}$
\cite{CMS}. In what follows, we will carry out calculations of
cross-sections of this process for the center of mass energy
$14\,\textrm{TeV}$ and the masses of KK resonances larger than
this exclusion limit.

The process is mainly due to two partonic processes $u\bar u \to
\mu^+\mu^-$ and $d\bar d \to \mu^+\mu^-$, the first one giving the
main contribution. In the general case there are  three KK towers
that essentially contribute to these processes and their lowest
excitations may be present in the LHC energy range: $Z^{\prime}$,
$\gamma^{\prime}$ and the first massive tensor graviton
$gr^{\prime}$.  As we explained in the previous section, we
neglect the contribution from the interaction with the Higgs field
to the KK  masses of the vector field excitations  and take the
masses of the first two modes $\gamma'$ and $Z'$ to be the same
and equal to $M_{\gamma'}=M_{Z'}=5\,\textrm{TeV}$.  The
corresponding widths of the $\gamma'$ and  $Z'$  resonances have
been calculated and turned out to be $\Gamma_{\gamma'}=0.10
 \,\textrm{TeV}$ and $\Gamma_{Z'}=0.15  \,\textrm{TeV}$
respectively.

As we explained in the previous section, the mass of the graviton
excitation should be taken approximately equal to
$M_{gr^{\prime}}=7.5\,\textrm{TeV}$ in this case, and its
(inverse) coupling constant is chosen to be $\Lambda_\pi=
10\,\textrm{TeV}$, whereas the width
$\Gamma_{gr^{\prime}}=0.41\,\textrm{TeV}$ has been calculated with
the help of the formulas of paper \cite{BBSV}.

The effective Lagrangians for the KK towers of $\gamma$ and $Z$
have the same form as (\ref{W_KK}) with the weak charged current
$J^{+\mu } $ replaced by the electromagnetic current and the weak
neutral current respectively. In this case the process $u\bar u
\to \mu^+\mu^-$ is described in
 our approximation by the following diagrams:

\vspace*{1cm}
\begin{picture}(93,81)(0,0)
\ArrowLine(13.5,78.0)(40.5,64.5) \Text(12.5,78.0)[r]{$u$}
\ArrowLine(40.5,64.5)(13.5,51.0) \Text(12.5,51.0)[r]{$\bar u$}
%\DashArrowLine(40.5,64.5)(67.5,64.5){3.0}
%\DashArrowLine(40.5,64.5)(67.5,64.5){3.0}
\Photon(40.5,64.5)(67.5,64.5){3}{3.0} \Text(54.8,70.5)[b]{$\gamma
$} \ArrowLine(94.5,78.0)(67.5,64.5) \Text(95.5,78.0)[l]{$\mu^+$}
\ArrowLine(67.5,64.5)(94.5,51.0) \Text(95.5,51.0)[l]{$\mu^-$}
\Text(116.8,60.5)[b]{\begin{Large}$+$\end{Large}}
%\Text(46,0)[b] {diagr.1}
\end{picture} \
\vspace{5mm} {} \qquad\allowbreak
\begin{picture}(93,81)(0,0)
\ArrowLine(13.5,78.0)(40.5,64.5) \Text(12.5,78.0)[r]{$u$}
\ArrowLine(40.5,64.5)(13.5,51.0) \Text(12.5,51.0)[r]{$\bar u$}
\Photon(40.5,64.5)(67.5,64.5){3}{3.0} \Text(54.8,70.5)[b]{$Z$}
\ArrowLine(94.5,78.0)(67.5,64.5) \Text(96.5,78.0)[l]{$\mu^+$}
\ArrowLine(67.5,64.5)(94.5,51.0) \Text(96.5,51.0)[l]{$\mu^-$}
\Text(116.8,60.5)[b]{\begin{Large}$+$\end{Large}}
%\Text(46,0)[b] {diagr.2}
\end{picture} \
\vspace{5mm} {} \qquad\allowbreak
%  diagram # 2
\begin{picture}(93,81)(0,0)
\ArrowLine(13.5,78.0)(40.5,64.5) \Text(12.5,78.0)[r]{$u$}
\ArrowLine(40.5,64.5)(13.5,51.0) \Text(12.5,51.0)[r]{$\bar u$}
%\DashArrowLine(40.5,64.5)(67.5,64.5){3.0}
%\DashArrowLine(40.5,64.5)(67.5,64.5){3.0}
\Photon(40.5,64.5)(67.5,64.5){3}{3.0}
\Text(54.8,70.5)[b]{$\gamma^{\prime}$}
\ArrowLine(94.5,78.0)(67.5,64.5) \Text(95.5,78.0)[l]{$\mu^+$}
\ArrowLine(67.5,64.5)(94.5,51.0) \Text(95.5,51.0)[l]{$\mu^-$}
\Text(116.8,60.5)[b]{\begin{Large}$+$\end{Large}}
%\Text(46,0)[b] {diagr.2}
\end{picture} \
%  diagram # 3

\vspace*{-2cm}
\begin{picture}(93,81)(0,0)
\ArrowLine(13.5,78.0)(40.5,64.5) \Text(12.5,78.0)[r]{$u$}
\ArrowLine(40.5,64.5)(13.5,51.0) \Text(12.5,51.0)[r]{$\bar u$}
%\DashArrowLine(40.5,64.5)(67.5,64.5){3.0}
%\DashArrowLine(40.5,64.5)(67.5,64.5){3.0}
\Photon(40.5,64.5)(67.5,64.5){3}{3.0}
\Text(54.8,70.5)[b]{$Z^{\prime}$} \ArrowLine(94.5,78.0)(67.5,64.5)
\Text(95.5,78.0)[l]{$\mu^+$} \ArrowLine(67.5,64.5)(94.5,51.0)
\Text(95.5,51.0)[l]{$\mu^-$}
\Text(116.8,60.5)[b]{\begin{Large}$+$\end{Large}}
%\Text(46,0)[b] {diagr.1}
\end{picture} \
\vspace{5mm} {} \qquad\allowbreak
\begin{picture}(93,81)(0,0)
\ArrowLine(13.5,78.0)(30.5,64.5) \Text(12.5,78.0)[r]{$u$}
\ArrowLine(30.5,64.5)(13.5,51.0) \Text(12.5,51.0)[r]{$\bar u$}
%\DashArrowLine(40.5,64.5)(67.5,64.5){3.0}
%\Text(54.8,67.5)[b]{$A+$}
%\ArrowLine(94.5,78.0)(67.5,64.5)
\Text(24.0,76.0)[l]{{\scriptsize  C$\gamma$}}
\Text(27.0,64.0)[l]{\begin{large}$ \bullet$\end{large}}
\ArrowLine(47.5,78.0)(30.5,64.5) \Text(49.0,78.0)[l]{$\mu^+$}
\ArrowLine(30.5,64.5)(47.5,51.0) \Text(49.0,51.0)[l]{$\mu^-$}
\Text(94.8,60.5)[b]{\begin{Large}$+$\end{Large}}
%\Text(46,0)[b] {diagr.5}
\end{picture} \
\vspace{5mm} {} \qquad\allowbreak
%  diagram # 2
\begin{picture}(93,81)(0,0)
\ArrowLine(13.5,78.0)(30.5,64.5) \Text(12.5,78.0)[r]{$u$}
\ArrowLine(30.5,64.5)(13.5,51.0) \Text(12.5,51.0)[r]{$\bar u$}
%\DashArrowLine(40.5,64.5)(67.5,64.5){3.0}
%\Text(54.8,67.5)[b]{$A+$}
%\ArrowLine(94.5,78.0)(67.5,64.5)
\Text(25.0,76.0)[l]{{\scriptsize  CZ}}
\Text(27.0,64.0)[l]{\begin{large}$ \bullet$\end{large}}
\ArrowLine(47.5,78.0)(30.5,64.5) \Text(49.0,78.0)[l]{$\mu^+$}
\ArrowLine(30.5,64.5)(47.5,51.0) \Text(49.0,51.0)[l]{$\mu^-$}
\Text(96.8,60.5)[b]{\begin{Large}$+$\end{Large}}
%\Text(162.8,60.5)[b]{(15)}
%\Text(46,0)[b] {diagr.6}
%\Text(165.8,60.5)[b]{\addtocounter{equation}{1}(\arabic{equation})}
\end{picture}
%  diagram # 3

\vspace*{-2cm}
%\vspace{5mm} {} \qquad\allowbreak \qquad\allowbreak
%  diagram # 3
\begin{picture}(93,81)(0,0)
\ArrowLine(13.5,78.0)(40.5,64.5) \Text(12.5,78.0)[r]{$u$}
\ArrowLine(40.5,64.5)(13.5,51.0) \Text(12.5,51.0)[r]{$\bar u$}
\Photon(40.5,64.5)(67.5,64.5){3}{3.0}
\Text(54.8,70.5)[b]{$gr^{\prime}$}
\ArrowLine(94.5,78.0)(67.5,64.5) \Text(96.5,78.0)[l]{$\mu^+$}
\ArrowLine(67.5,64.5)(94.5,51.0) \Text(96.5,51.0)[l]{$\mu^-$}
\Text(116.8,60.5)[b]{\begin{Large}$+$\end{Large}}
%\Text(46,0)[b] {diagr.2}
\end{picture} \
\vspace{5mm} {} \qquad\allowbreak
%  diagram # 2
\begin{picture}(93,81)(0,0)
\ArrowLine(13.5,78.0)(30.5,64.5) \Text(12.5,78.0)[r]{$u$}
\ArrowLine(30.5,64.5)(13.5,51.0) \Text(12.5,51.0)[r]{$\bar u$}
%\DashArrowLine(40.5,64.5)(67.5,64.5){3.0}
%\Text(54.8,67.5)[b]{$A+$}
%\ArrowLine(94.5,78.0)(67.5,64.5)
\Text(25.0,76.0)[l]{{\scriptsize  Cgr}}
\Text(27.0,64.0)[l]{\begin{large}$ \bullet$\end{large}}
\ArrowLine(47.5,78.0)(30.5,64.5) \Text(49.0,78.0)[l]{$\mu^+$}
\ArrowLine(30.5,64.5)(47.5,51.0) \Text(49.0,51.0)[l]{$\mu^-$}
%\Text(162.8,60.5)[b]{(15)}
%\Text(46,0)[b] {diagr.6}
\Text(315.8,60.5)[b]{\addtocounter{equation}{1}(\arabic{equation})}
\setcounter{diags1}{\value{equation}}
\end{picture}

\vspace*{-2cm} \noindent  For simplicity,  we will again assume
that the coupling constants of the lowest excitations are the same
as those of the zero modes. Then the contact term C$\gamma$ is a
local product of two electromagnetic currents with the effective
coupling constant $e^2/M^2_{\gamma'\_sum}$,
 and the contact term CZ is a local product of two weak neutral
 currents  with the effective coupling constant
 $g^2 (cos(\theta_W))^{-2}/M^2_{Z'\_sum}$; for the chosen masses
of ${\gamma'}$ and ${Z'}$ boson the effective masses  turn out  to
be $M_{\gamma'\_sum}= M_{Z'\_sum}= 7 \,\textrm{TeV}$. The contact
term Cgr is given by formula (\ref{tensor_intCRS}), from which the
contribution of the first graviton excitation has been subtracted,
i.e., with the coupling constant
${(C-1)}/({\Lambda_{\pi}^{2}M_{gr^{\prime}}^{2}})$. In the
stabilized Randall-Sundrum model we approximately have $C \approx
1.8$ \cite{BBSV}, in UED models with the flat extra dimension the
constant turns out to be $C \approx 1.5$. In our subsequent
calculations we will assume $C=1.7$. Of course, there are the same
diagrams with the up quark replaced by the down quark. The
structure of the amplitude corresponding to diagrams
(\arabic{equation})  is much more complicated, than in the case of
the $W^{\prime}$ boson (\ref{amp}).

\begin{figure*}[!h!]
\begin{center}

\begin{minipage}[t]{.45\linewidth}
%\begin{minipage}[t][2cm][t]{0.4\textwidth}
\centering
\includegraphics[width=80mm,height=70mm]{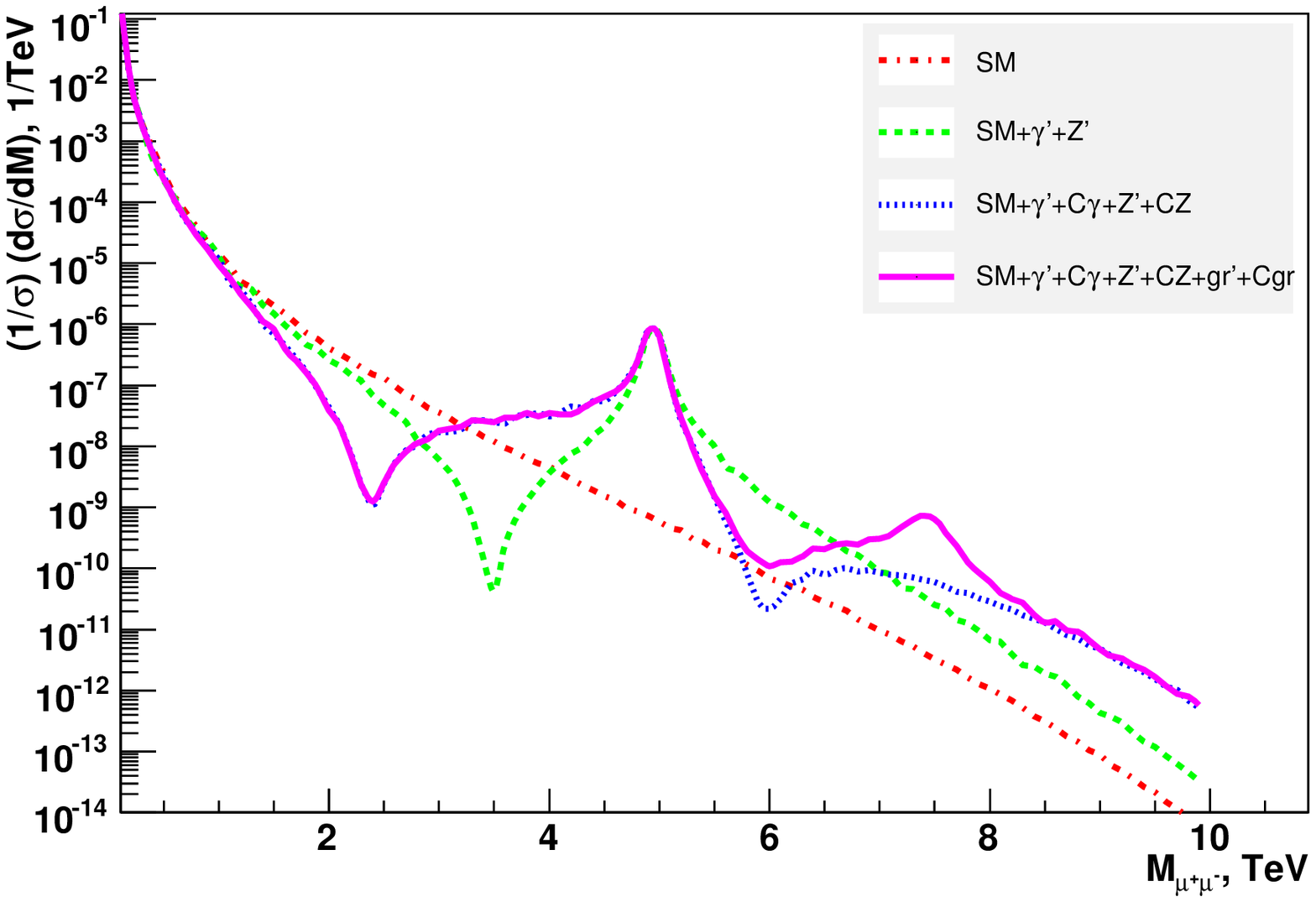}
\caption{\label{pic3} \footnotesize Invariant mass distribution for the Drell-Yan
process at the LHC with the center of mass energy 14~TeV   for
$M_{\gamma'}=M_{Z'}=5\,\textrm{TeV}$,
$\Gamma_{\gamma'}=0.10\,\textrm{TeV}$,
 $\Gamma_{Z'}=0.15\,\textrm{TeV}$,
 $M_{gr'} = 7.5\,\textrm{TeV} $, $\Gamma_{gr'}=0.41\,\textrm{TeV}$. }
\end{minipage}
\hfill
\begin{minipage}[t]{.45\linewidth}
\centering
\includegraphics[width=80mm,height=70mm]{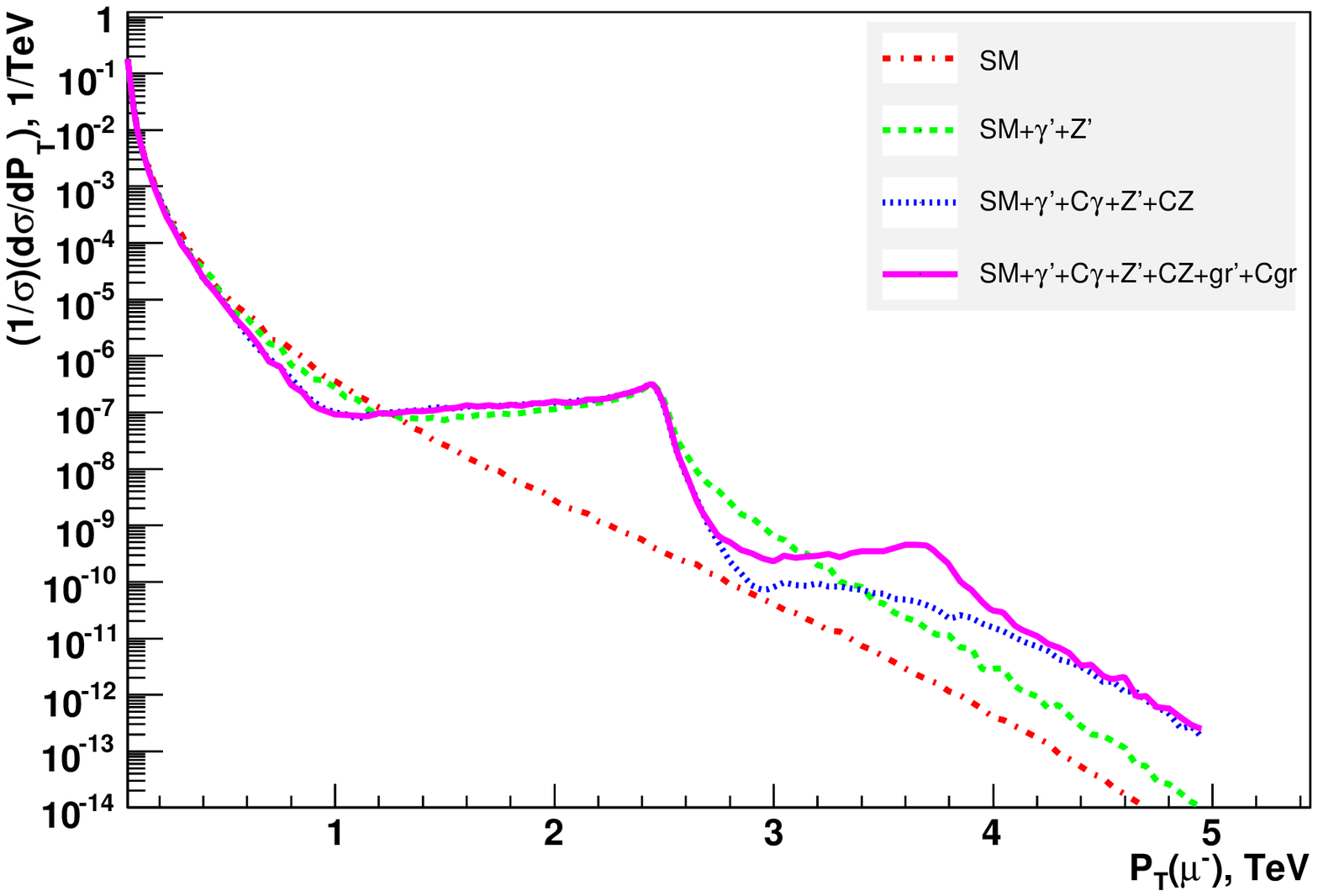}
\caption{ \label{pic4} \footnotesize  $P_{T}$ distribution for the Drell-Yan
process at the LHC with the center of mass energy 14~TeV   for
$M_{\gamma'}=M_{Z'}=5\,\textrm{TeV}$,
$\Gamma_{\gamma'}=0.10\,\textrm{TeV}$,
 $\Gamma_{Z'}=0.15\,\textrm{TeV}$,
 $M_{gr'} = 7.5\,\textrm{TeV} $, $\Gamma_{gr'}=0.41\,\textrm{TeV}$.}
\end{minipage}

\end{center}
\end{figure*}

We have calculated the cross-sections of the process $pp \to
\mu^+\mu^- + X$  for the sum of the first two diagrams (SM), for
the sum of the first four diagrams (SM + $\gamma^{\prime}$ +
$Z^{\prime}$) and for the sum of the first six diagrams (SM +
$\gamma^{\prime}$ + $Z^{\prime}$ + C$\gamma$ + CZ). The
corresponding distributions are plotted in figures \ref{pic3} and
\ref{pic4} (such distributions for smaller  and unequal masses of
$M_{\gamma'}$ and $M_{Z'}$ have been obtained in
\cite{Boos:2011ib}). It is clear that the interference with the
contact interaction terms is quite definite and is  very similar
to that in the single top production.

%=========================================================================
%=============================================================================

The cross-sections of the process $pp \to \mu^+\mu^- + X$ taking
into account all the diagrams  were calculated as well and are
also shown in figures \ref{pic3} and \ref{pic4}. In these
calculations the diagrams of gluon-gluon fusion to the excitations
of the graviton, which are important at the LHC and are shown
below, were also taken into account.

\vspace*{1cm}
%\vspace{5mm} {} \qquad\allowbreak \qquad\allowbreak
%  diagram # 3
\begin{picture}(93,81)(0,0)
\Photon(13.5,78.0)(40.5,64.5){2}{4.0} \Text(12.5,78.0)[r]{$g$}
\Photon(40.5,64.5)(13.5,51.0){2}{4.0} \Text(12.5,51.0)[r]{$g$}
\Photon(40.5,64.5)(67.5,64.5){3}{3.0}
\Text(54.8,70.5)[b]{$gr^{\prime}$}
\ArrowLine(94.5,78.0)(67.5,64.5) \Text(96.5,78.0)[l]{$\mu^+$}
\ArrowLine(67.5,64.5)(94.5,51.0) \Text(96.5,51.0)[l]{$\mu^-$}
\Text(116.8,60.5)[b]{\begin{Large}$+$\end{Large}}
%\Text(46,0)[b] {diagr.2}
\end{picture} \
\vspace{5mm} {} \qquad\allowbreak
%  diagram # 2
\begin{picture}(93,81)(0,0)
\Photon(13.5,78.0)(30.5,64.5){2}{4.0} \Text(12.5,78.0)[r]{$g$}
\Photon(30.5,64.5)(13.5,51.0){2}{4.0} \Text(12.5,51.0)[r]{$g$}
%\DashArrowLine(40.5,64.5)(67.5,64.5){3.0}
%\Text(54.8,67.5)[b]{$A+$}
%\ArrowLine(94.5,78.0)(67.5,64.5)
\Text(25.0,76.0)[l]{{\scriptsize  Cgr}}
\Text(27.0,64.0)[l]{\begin{large}$ \bullet$\end{large}}
\ArrowLine(47.5,78.0)(30.5,64.5) \Text(49.0,78.0)[l]{$\mu^+$}
\ArrowLine(30.5,64.5)(47.5,51.0) \Text(49.0,51.0)[l]{$\mu^-$}
%\Text(162.8,60.5)[b]{(15)}
%\Text(46,0)[b] {diagr.6}
\Text(315.8,60.5)[b]{\addtocounter{equation}{1}(\arabic{equation})}
\end{picture}
\vspace*{-1cm}

\begin{figure*}[!h!]
\begin{center}

\begin{minipage}[t]{.45\linewidth}
%\begin{minipage}[t][2cm][t]{0.4\textwidth}
\centering
\includegraphics[width=80mm,height=70mm]{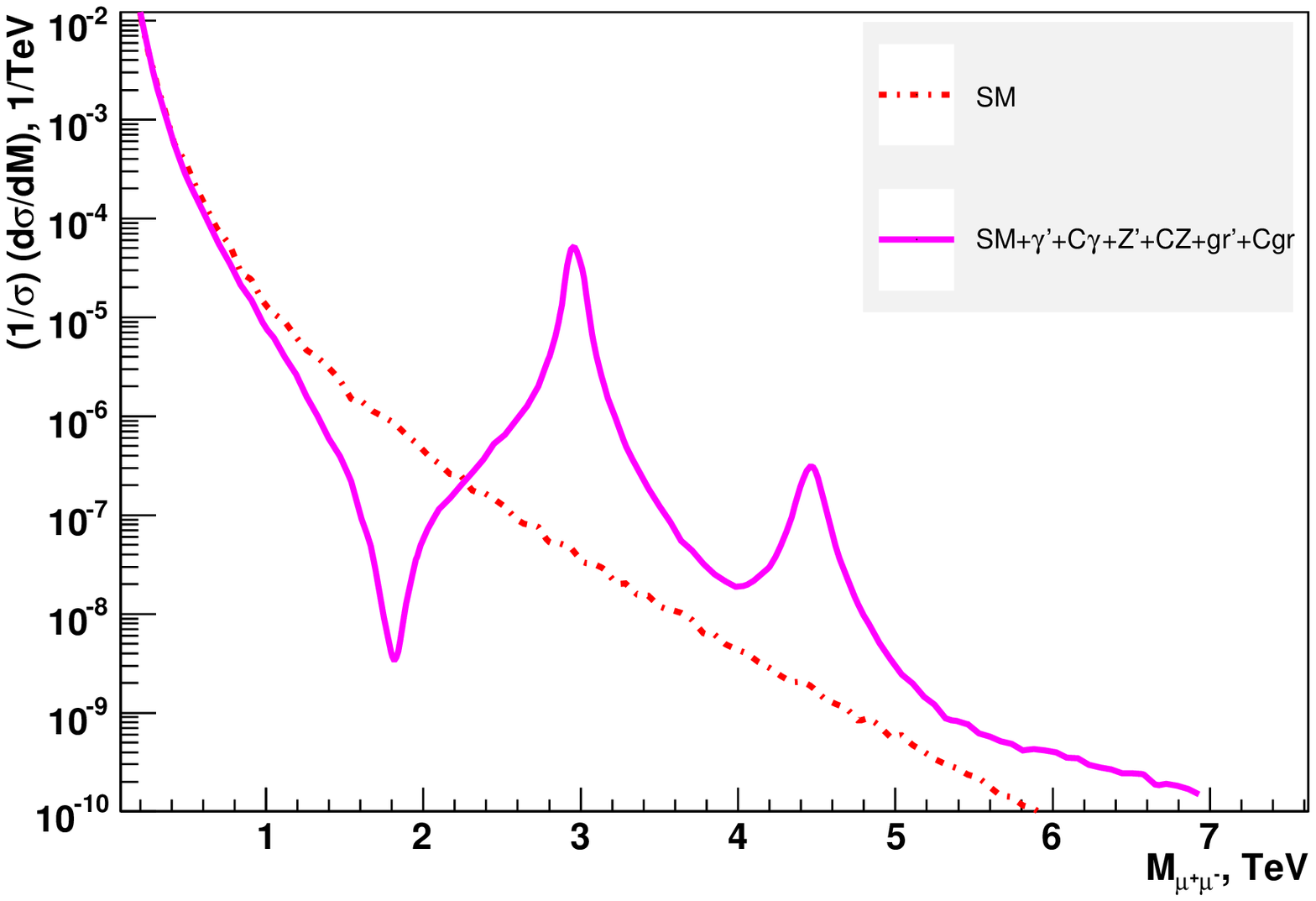}
\caption{\label{pic5} \footnotesize Invariant mass distribution for the Drell-Yan
process at the LHC with the center of mass energy 14~TeV   for
$M_{\gamma'}= M_{Z'}= 3\,\textrm{TeV}$,
$\Gamma_{\gamma'}=0.06\,\textrm{TeV}$,
 $\Gamma_{Z'}=0.09\,\textrm{TeV}$,
 $M_{gr'} = 4.5\,\textrm{TeV} $, $\Gamma_{gr'}=0.14\,\textrm{TeV}$.}
\end{minipage}
\hfill
\begin{minipage}[t]{.45\linewidth}
\centering
\includegraphics[width=80mm,height=70mm]{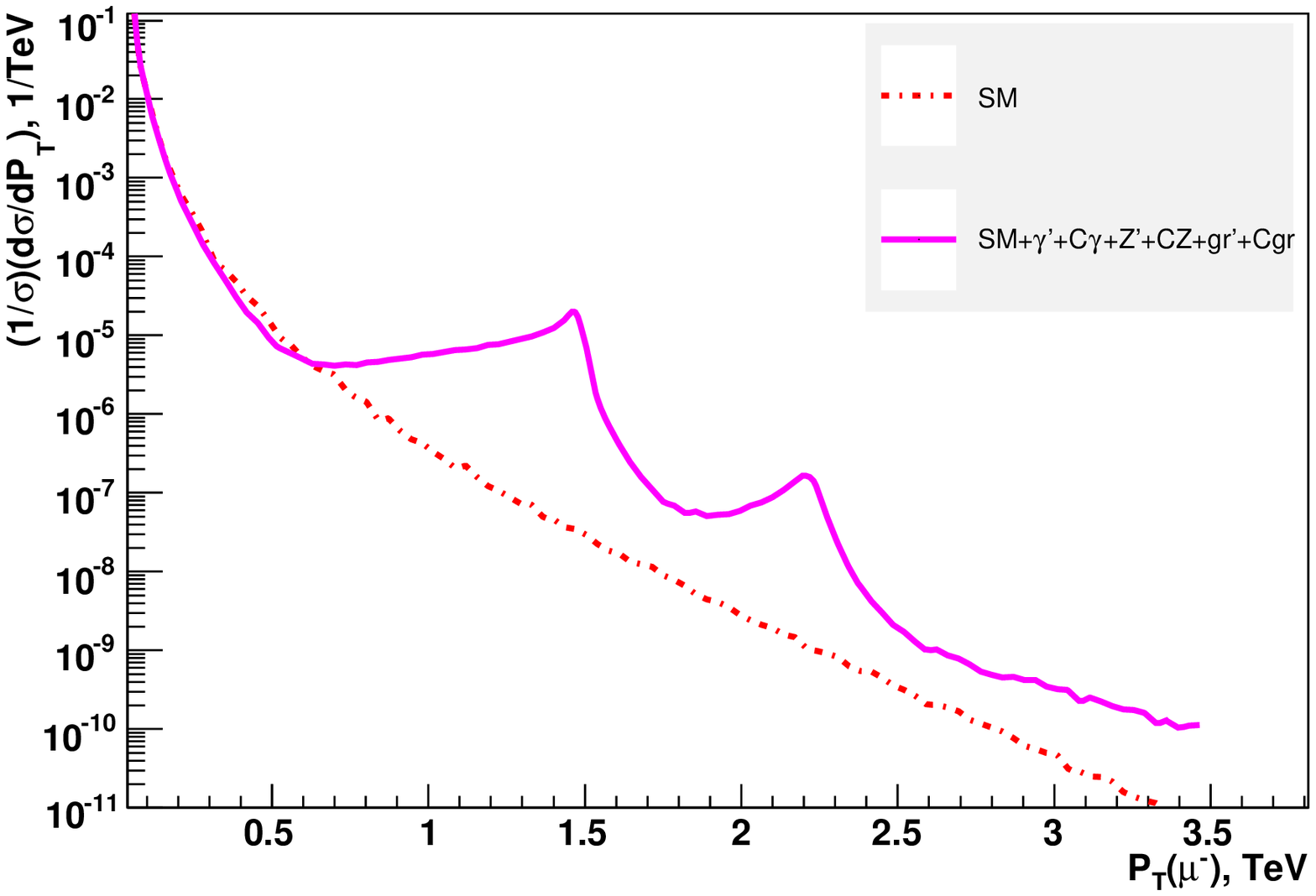}
\caption{ \label{pic6} \footnotesize  $P_{T}$ distribution for the Drell-Yan
process at the LHC with the center of mass energy 14~TeV   for
$M_{\gamma'}= M_{Z'}= 3\,\textrm{TeV}$,
$\Gamma_{\gamma'}=0.06\,\textrm{TeV}$,
 $\Gamma_{Z'}=0.09\,\textrm{TeV}$,
 $M_{gr'} = 4.5\,\textrm{TeV} $, $\Gamma_{gr'}=0.14\,\textrm{TeV}$.}
\end{minipage}

\end{center}
\end{figure*}

\begin{figure*}[!h!]
\begin{center}

\begin{minipage}[t]{.45\linewidth}
%\begin{minipage}[t][2cm][t]{0.4\textwidth}
\centering
\includegraphics[width=80mm,height=70mm]{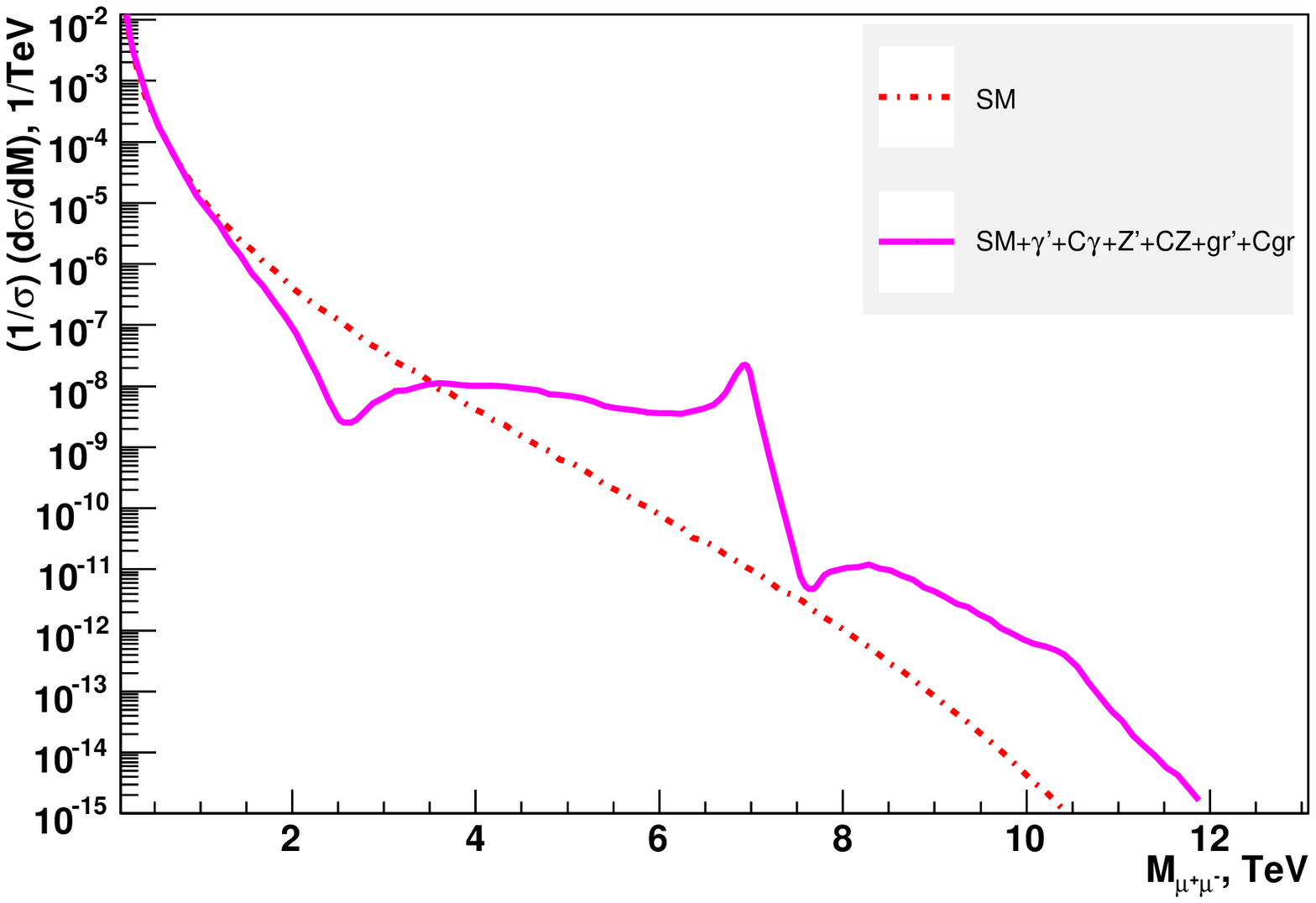}
\caption{\label{pic7} \footnotesize Invariant mass distribution for the Drell-Yan
process at the LHC with the center of mass energy 14~TeV   for
$M_{\gamma'}= M_{Z'}= 7\,\textrm{TeV}$,
$\Gamma_{\gamma'}=0.15\,\textrm{TeV}$,
 $\Gamma_{Z'}=0.22\,\textrm{TeV}$,
 $M_{gr'} = 10.5\,\textrm{TeV} $, $\Gamma_{gr'}=0.57\,\textrm{TeV}$.}
\end{minipage}
\hfill
\begin{minipage}[t]{.45\linewidth}
\centering
\includegraphics[width=80mm,height=70mm]{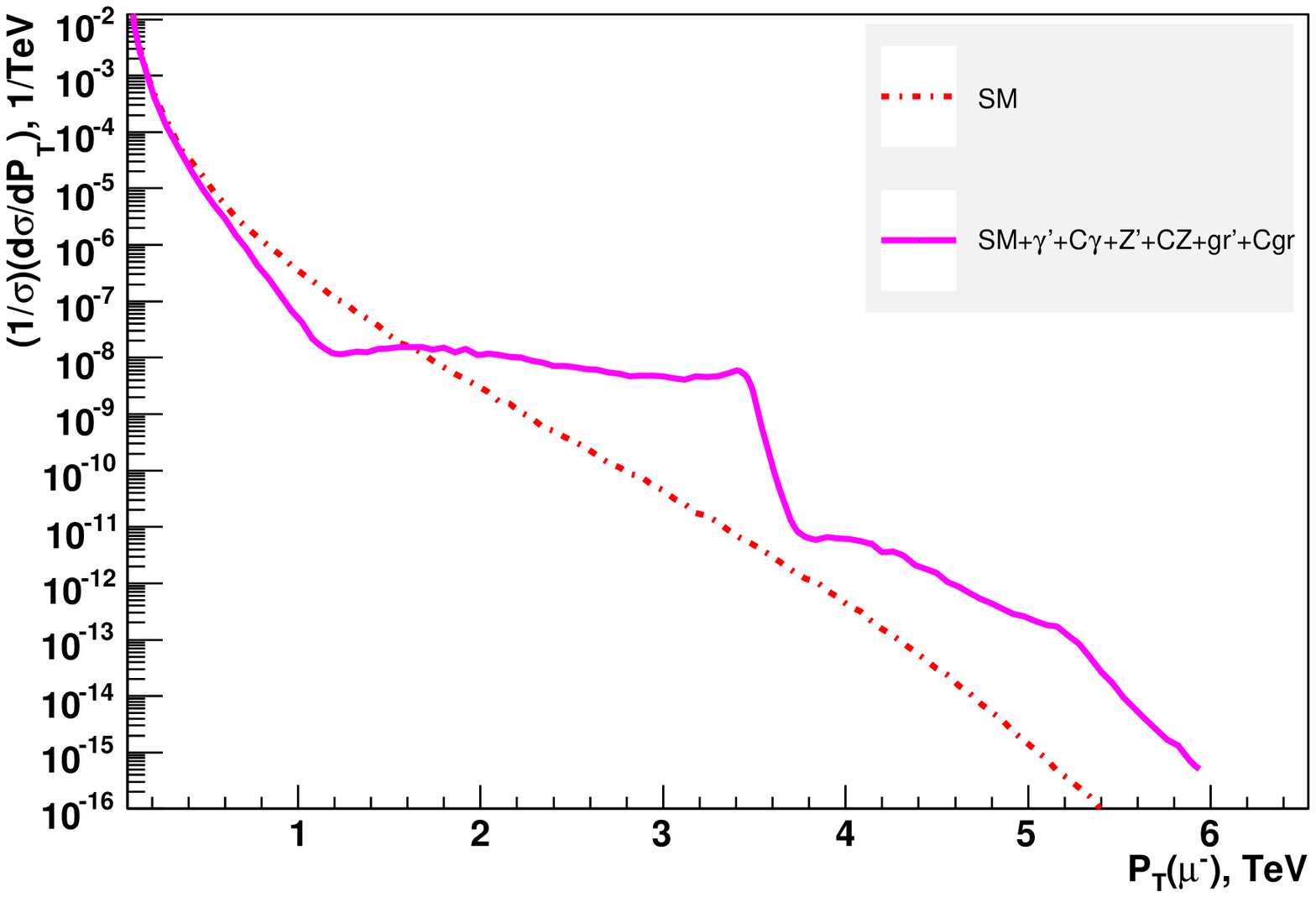}
\caption{ \label{pic8} \footnotesize $P_{T}$ distribution for the Drell-Yan
process at the LHC with the center of mass energy 14~TeV   for
$M_{\gamma'}= M_{Z'}= 7\,\textrm{TeV}$,
$\Gamma_{\gamma'}=0.15\,\textrm{TeV}$,
 $\Gamma_{Z'}=0.22\,\textrm{TeV}$,
 $M_{gr'} = 10.5\,\textrm{TeV} $, $\Gamma_{gr'}=0.57\,\textrm{TeV}$.}
\end{minipage}

\end{center}
\end{figure*}

The calculations of the cross-sections of the process $pp \to
\mu^+\mu^- + X$ taking into account all the diagrams
(\arabic{diags1}) and (\arabic{equation}) were carried out for two
more sets of excitation masses. The results for the set $
M_{\gamma'} =  M_{Z'} = 3\,\textrm{TeV}, M_{gr'} = 4.5
\,\textrm{TeV}$ ($M_{\gamma'\_sum}= M_{Z'\_sum}= 4.2
\,\textrm{TeV},
\,\Gamma_{\gamma^{\prime}}=0.06\,\textrm{TeV},\,\,\Gamma_{Z^{\prime}}=0.09\,\textrm{TeV},\,
\Gamma_{gr^{\prime}}=0.14\,\textrm{TeV},\,\Lambda_{\pi}=
8\,\textrm{TeV} $) are shown in figures \ref{pic5} and \ref{pic6},
and those for the set $M_{\gamma'} = M_{Z'} = 7\,\textrm{TeV},
M_{gr'}     = 10.5 \,\textrm{TeV} $ ($M_{\gamma'\_sum}=
M_{Z'\_sum}= 9.8 \,\textrm{TeV},
\,\Gamma_{\gamma^{\prime}}=0.15\,\textrm{TeV},\,\,\Gamma_{Z^{\prime}}=0.22\,\textrm{TeV},
\,\Gamma_{gr^{\prime}}=0.57\,\textrm{TeV},
\,\Lambda_{\pi}=14\,\textrm{TeV}$) are shown in figures
\ref{pic7} and \ref{pic8}.

%=========================================================================
%=============================================================================

%=========================================================================
%=============================================================================

An interesting feature of all these plots is the absence of an
interference between the first graviton KK excitation and that of
the Z boson (the interference with the photon and its excitations
is forbidden in QFT). One can rigorously prove that there is no
interference in the distributions in the invariant mass of the
$\mu^+ \mu^-$-pair, because the interference terms vanish after
the integration over the angles. Explicit calculations show that
the interference is also  equal to zero for the distributions in
the transverse momentum.

\section{Concluding remarks}

At this point we have to make several remarks. First, as it was
noted in \cite{Boos:2011ib}, there is a good reason to believe
that the interference picture discussed here  is not destroyed by
the NLO corrections. The pole structure of the amplitude, which
leads to the non-trivial interference, is clearly not affected by
the QCD corrections to the external lines and to the vertices. The
most dangerous terms seem to be those of self-energy diagrams, but
they are defined so as to vanish on the mass shell and therefore
contribute only to the particle widths and to the renormalization
of mass.

Second, the presence of just $W''$ or $Z''$ or $\gamma''$ can
produce similar effects. But if such effects are found, it is easy
to improve our approach by taking into account the contributions
of $W''$, $Z''$  and $\gamma''$ exactly and approximating the
contributions of the towers above them by contact interactions
(however, in order to handle such a situation the high luminosity
and high energy regime of the LHC operation would be needed). As
we have seen, the contribution to the amplitude of the tower above
a mode is of the same order as the contribution of the mode proper
away from the resonance. Thus, on the basis of our analysis one
can think that there should be distinct differences between the KK
scenario and that with only two excitations of the SM electroweak
gauge bosons. Therefore, an observation of such interference
effects for $W'$, $Z'$, $\gamma'$ and the presence of the first
tensor KK graviton can be interpreted as a strong argument in
favor of the existence of extra dimensions.

Third, we would like to note once again that the masses of $W'$,
$Z'$, and $\gamma'$ resonances are nearly degenerate, and
therefore one should  observe resonances with practically the same
masses in the single top  and $\mu^+ \nu_\mu$ production processes
and Drell-Yan processes, which is an interesting prediction of
models with extra dimensions.

Finally, we would also like  to mention here that the scenario
discussed in the present paper is very close to the framework of
warped extra dimension with the Standard Model fields propagating
in it \cite{Agashe:2013fda}, if one considers only the processes,
to which the excitations of the SM fermions do not contribute.

Thus, our analysis shows that in order to search correctly for the
Kaluza-Klein excitations of SM particles, in particular, to put
correctly exclusion limits, it is necessary, in modeling the
signal, to sum the contributions to the amplitudes of all the KK
modes above the resonances and to take into account their
interference with the contribution of the resonances. Our
calculations of the single top and $\mu^{+}\nu_{\mu}$ production
and Drell-Yan processes also show that the lowest excitations of
the SM gauge bosons, as well as that of the graviton, may be in
principle observed at the LHC with the center of mass energy  $14
\,\textrm{TeV}$ if their masses are below $10 \,\textrm{TeV}$.
However, realistic simulations taking into account the
backgrounds are needed in order to find the LHC collider potential
in searches for KK states including  the interference with the
rest of KK towers.

\bigskip

{ \large \bf Acknowledgments}

The authors are grateful to  L. Dudko for useful discussions. The
work was supported in part by RFBR grants 13-02-01050,
12-02-93108-CNRSL-a and the grant of Russian Ministry of Education
and Science NS-3042.2014.2.

\end{document}